# Potential reversible hydrogen storage in Li-decorated carbon allotrope PAI-Graphene: A first-principles study

*Vikram Mahamiya[a*], Alok Shukla[a], Brahmananda Chakraborty[b,c]*

[a]Indian Institute of Technology Bombay, Mumbai 400076, India

[b]High pressure and Synchrotron Radiation Physics Division, Bhabha Atomic Research Centre, Bombay, Mumbai, India-40085

[c]Homi Bhabha National Institute, Mumbai, India-400094

email: vikram.physics@iitb.ac.in ; shukla@phy.iitb.ac.in ; brahma@barc.gov.in

## Abstract

Two-dimensional porous carbon nanomaterials are proven to be promising hydrogen storage substrates as they possess high surface area, large number of active sites, low molecular mass, and hydrogen molecules can be adsorbed on both sides of these materials. By performing first-principles density functional theory-based calculations, we report ultrahigh reversible hydrogen uptake in lithium decorated 2D carbon allotrope PAI-graphene, which is formed of a regular pattern of polymerized as-indacenes (PAI). We found that a single unit cell of PAI-graphene can be decorated by 8 Li atoms, in which each Li atom can reversibly adsorb 4 hydrogen molecules, leading to 15.7 % of H uptake, remarkably higher than the DOE demand of 6.5 %. Li atom donates its valence 2s-electron to PAI-graphene and gets ionized. The adsorption energies of the various $H_2$ attached to Li-atom are found to be suitable for reversible use during practical applications. Hydrogen molecules get attached to the ionized metal atom by electrostatic interactions. An energy barrier of 1.48 eV is present for the diffusion of Li atoms between the two most stable adsorption sites which justifies the absence of the clustering of Li atoms.

# Introduction

Alternative energy demands are increasing globally due to the sharp rise in population as well as in living standards. Hydrogen energy can replace the fossil fuel dependence in automotive applications since hydrogen is highly abundant and possesses a large energy density of 120 MJ/Kg[1,2]. The combustion of fossil fuel produces greenhouse gases as byproducts, but hydrogen combustion is environmentally friendly. Hydrogen production, storage, and distribution are the real challenges for researchers in present times. Hydrogen can be produced by various renewable methods, including sunlight, wind energy, biomass, thermal energy, etc., through the electrolysis of water. However, half of the global hydrogen is produced through the steam reformation of natural gases[3]. Safe, economical, and compact hydrogen storage is required for practical applications, and therefore, the gas and liquid form of hydrogen is not suitable since it is costly and involves safety concerns[4]. Solid-phase of hydrogen is desirable in this regard, where hydrogen is stored in the form of chemical hydrides. According to the department of energy, United-States (DOE-US) hydrogen storage guidelines, the host material should hold at least 6.5 wt % of H, and the binding energy (B.E.) of the H molecules should lie in the -0.1 eV/$H_2$ to -0.7 eV/$H_2$ range [5,6]. People have explored metal alloys and hydrides[7–13], Porous zeolites and MOFs[14–18], and various carbon nanostructures, including 0D fullerenes[19–27], 1D carbon nanotubes[28–37], 2D graphene, and graphynes[38–50], holey graphyne[51,52], biphenylene sheet [53,54], etc., for hydrogen storage. 2D carbon nanomaterials are proven to be promising hosts since they possess relatively lower molecular mass, large surface area, and hydrogen can be adsorbed to the front as well as to the backside of nanomaterials. Pristine carbon nanostructures bind hydrogen molecules only through van der Waals interactions and, therefore, they are inefficient for reversible hydrogen storage applications under ambient conditions [55,56,57]. The binding of hydrogen molecules improves when nanostructures are decorated with metal atoms since strong electrostatic and Kubas interactions [58] also play a vital role in hydrogen adsorption to these structures. Generally, transition metals (TMs) decorated hosts bind more $H_2$ compared to alkali or alkali-earth metal (AMs or AEMs) decorated hosts due to the presence of strong Kubas interactions [59], but the problem is that the cohesive energies of TMs are significantly large [60], and therefore, they tend to make clusters instead of single site adsorption which reduces the hydrogen uptake drastically. The cohesive energy of AMs and AEMs is lower than the TMs [61–63], which reduces the possibility of metal-metal clustering. Li decorated systems are very promising hydrogen storage materials because of the lesser molecular weight and cohesive energy of Li compared to most of the metals. A hydrogen adsorption study in Li decorated $C_{60}$ fullerene structure was done by Sun et al.[61]. They have found that one molecule of $C_{60}$ fullerene can adsorb 12 Li atoms, and 60 $H_2$ molecules can be stored by Li decorated $C_{60}$ fullerene system. Sahoo and coworkers[27] reported reversible hydrogen uptake in alkali metal (Li,Na) decorated small fullerene $C_{20}$. They reported that the Li atom adsorbs five hydrogen molecules leading to a maximum hydrogen uptake of 13 %. Hydrogen storage in Li decorated graphene with vacancies is investigated by Seenithurai et al.[64]. They have estimated a moderate hydrogen uptake of 7.2 wt % with a B.E. of -0.26 eV/$H_2$. Wang et al.[65] have reported around 11 % of H uptake in the Li-decorated porous graphene structure. A remarkable gravimetric capacity of 18.6 % is predicted in graphyne structure decorated with Li atoms by Guo and coworkers[66]. A detailed study of H-storage in various metal atoms AM, AEM(Li, Ca), TM(Sc, Ti) decorated graphyne was performed by Guo and coworkers[67]. They have found that Li decorated structure has a maximum H uptake of 18.6 %. Li and Na decorated graphdiyne structures have been explored for hydrogen adsorption purposes by Wang and coworkers[68] using first-principles and Monte Carlo methods. They have reported a maximum of 8.8 % of H uptake with Li decoration. Gao and coworkers[69] have explored the hydrogen adsorption in Li decorated $B_2O$ and reported around 10 % of hydrogen uptake, significantly more than DOE-US guidelines. H storage in triazine-based frameworks decorated with AMs/AEMs (Li,Na,Ca) was explored by Chen and coworkers[70]. They found a maximum hydrogen uptake of 12 % for Li decorated triazine-based frameworks. Hydrogen adsorption properties of Li decorated biphenylene have been explored by Denis et al. [53] by using the VDW-DF method. Recently Mahamiya et al. [54] have explored the reversible $H_2$ storage mechanism in K and Ca decorated biphenylene sheet by employing the first-principles methods. Gao and coworkers[51] have reported reversible hydrogen adsorption in Li attached holey graphyne (HGY) structure which has been recently synthesized. They have predicted around 13 % of hydrogen uptake for the HGY structure decorated with Li, which is remarkably high.

Recently, a novel 2D metallic carbon allotrope PAI-graphene was theoretically predicted by Chen et al.[71]. PAI-graphene structure is composed of five, six, and seven-membered carbon rings, and it is found to be more stable than previously predicted penta-graphene[72] and psi-graphene[73] structures. The electronic properties of PAI-graphene can be tuned by applying the strain, and therefore, it has promising applications in nanoelectronic devices. Recently

Cheng et al.[74] have reported that the PAI-graphene structure is a potential anode material for lithium and sodium-ion battery applications due to its outstanding storage capacity.

Here, we have explored the H storage properties of the Li functionalized PAI-graphene structure by employing DFT simulations. Li atom is strongly attached to PAI-graphene structure with a B.E. value of -2 eV. We have plotted the density of states (DOS), projected DOS (PDOS), and spatial charge density difference plots to depict the charge transfer phenomenon between Li atom and PAI-graphene. The $H_2$ molecules are attached to the Li-decorated PAI-graphene structure by electrostatic interactions and van der Waals interactions. We have investigated the clustering of the Li atoms by calculating the energy barrier for the diffusion of metal atoms, which indicates that the clustering of Li atoms should not take place in the Li-decorated PAI-graphene system.

## Computational details

We have carried out DFT simulations as implemented in the Vienna Ab Initio Simulation Package(VASP)[75–78] with PBE-GGA[79] exchange-correlation functional. A single unit cell of PAI-graphene having 24 carbons is considered, along with a vacuum of 20 Å between two consecutive periodic layers of PAI-graphene. For the Brillouin zone sampling, we select a *k*-grid of 5*5*1 kpoints in the Monkhorst-Pack. We have taken 500 eV of kinetic energy cutoff for the plane-wave basis expansion. The convergence criteria for the force and energy are considered to be 0.01 eV/Å and $10^{-5}$ eV, respectively. We performed the climbing-image nudged-elastic-band (CI-NEB)[80] calculations to calculate the energy barrier for the diffusion of the Li atoms.

## Results and discussions

### Adsorption of Li atom on PAI-graphene

We have taken a single unit cell of PAI-graphene structure for Li-decoration and then hydrogen adsorption purposes. The relaxed unit cell of PAI-graphene is presented in **Fig. 1(a).** Next, we have kept the Li atoms at various sites of PAI-graphene, and performed the geometry optimization calculations. The relaxed geometry of the Li-decorated PAI-graphene system when the decoration of Li is above to the center of the pentagon (pent), hexagon (hex), and heptagon (hept) of PAI-graphene are presented in **Fig. 1 (b, c, & d)**, respectively. The B.E. of the Li atom when it is decorated on the pent, hex, and hept sites of the PAI-graphene is -1.95 eV, -1.98 eV, and -2.00 eV, respectively, which are more than the cohesive energy of the Li-atom -1.63 eV/atom [81,82] indicating that the host PAI-graphene structure binds the Li atom strongly and the possibilities of metal clustering are negligible. The bond distance between the Li atom and the nearest carbon of PAI-graphene is 2.2 Å when Li is decorated at the pent and hex sites of the PAI-graphene while it is 2.3 Å when Li is decorated at the hept site of PAI-graphene structure. We found that the B.E. of the Li atom on the PAI-graphene and Li-C distance are almost same for all the sites (pent, hex, and hept), so PAI-graphene structure can be decorated with the Li atom at any of the sites and have potential applications in Li storage. The B.E. of the Li atom decorated on the PAI-graphene structure is calculated by the following equation:

$$E_b(Li) = E(PAI - gra + Li) - E(PAI - gra) - E(Li) \qquad (1)$$

Here, $E(PAI - gra + Li)$, $E(PAI - gra)$, and $E(Li)$ denote the total energy of the Li-decorated PAI-graphene, PAI-graphene, and isolated Li atom, respectively.

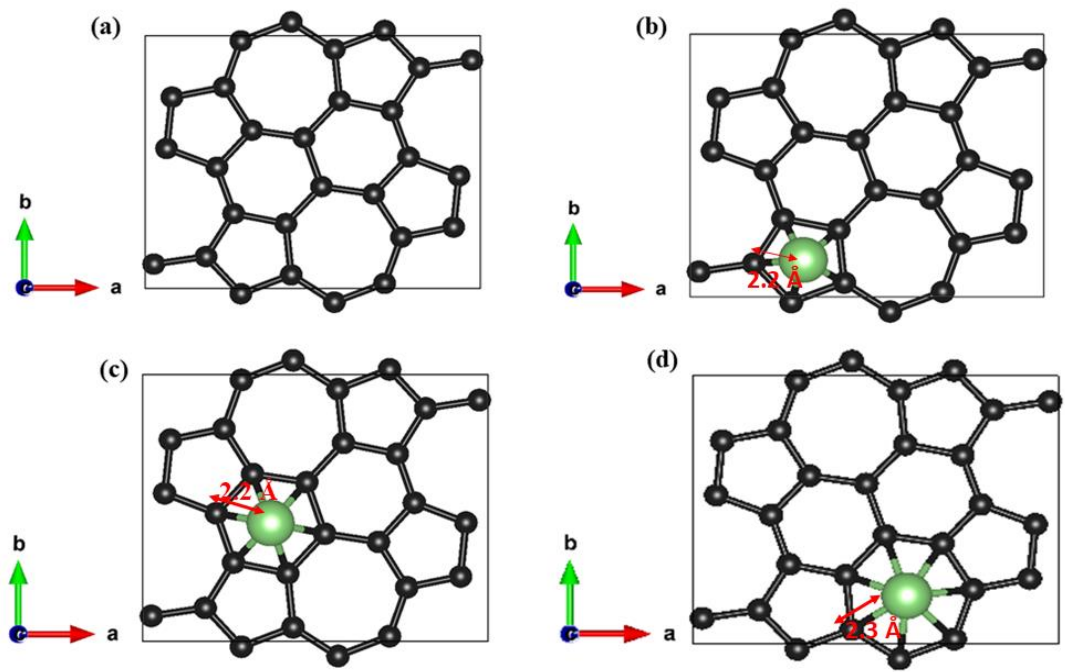

**Fig. 1. Relaxed geometries of (a) PAI-graphene (b) Li-decorated PAI-graphene where Li atom is decorated above to the center of pentagon (pent) (c) Li-decorated PAI-graphene**

where Li atom is decorated above to the center of hexagon (hex) **(d) Li-decorated PAI-graphene where Li atom is decorated above to the center of heptagon (hept). Here black and green color represent carbon and lithium atoms, respectively.**

**Orbital interactions and charge transfer between Li atom and PAI-graphene**

To investigate the electronic structure and orbital interaction, we present the total DOS of PAI-graphene and Li-decorated PAI-graphene systems in **Fig. 2(a)** and **Fig. 2(b)**, respectively. PAI-graphene is a metallic structure, as reported in the literature [71,74], which remains metallic after Li atom decoration. However, there is some enhancement in the DOS of PAI graphene after the decoration of the Li atom near the Fermi level, which is due to the interaction between the 2s-orbital of Li and 2p-orbitals of the carbon atoms.

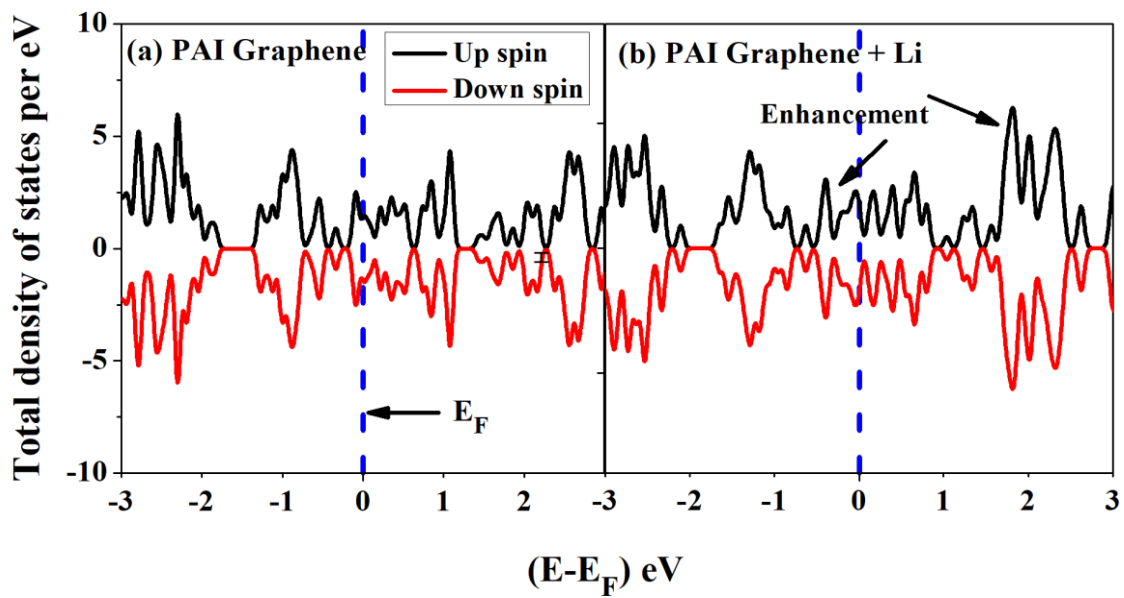

**Fig. 2. Total DOS of (a) PAI-graphene (b) Li-decorated PAI-graphene. Fermi level is adjusted at 0 eV.**

To get more clear insights regarding the orbital interactions, we present the PDOS of the C-2p orbitals of PAI-graphene before and after the adsorption of the Li-atom as shown in **Fig. 3 (a & b).** We observe more intense states for the C-2p orbitals in the case of Li-decorated PAI-graphene compared to pristine PAI-graphene near the Fermi level. This indicates that charge is getting transferred between Li valence shell orbitals to the valence orbitals of the carbon atoms of PAI-graphene. We also present the PDOS of the Li-2s orbital for isolated Li atom and Li-decorated PAI-graphene system presented in **Fig. 3 (c & d),** which clearly shows the states near the Fermi level of isolated Li, are absent in the case of Li-decorated PAI-graphene system. Therefore, a finite amount of charge flows from the Li-2s orbital to the C-2p orbitals of PAI-graphene, resulting in the strong binding of Li on PAI-graphene structure.

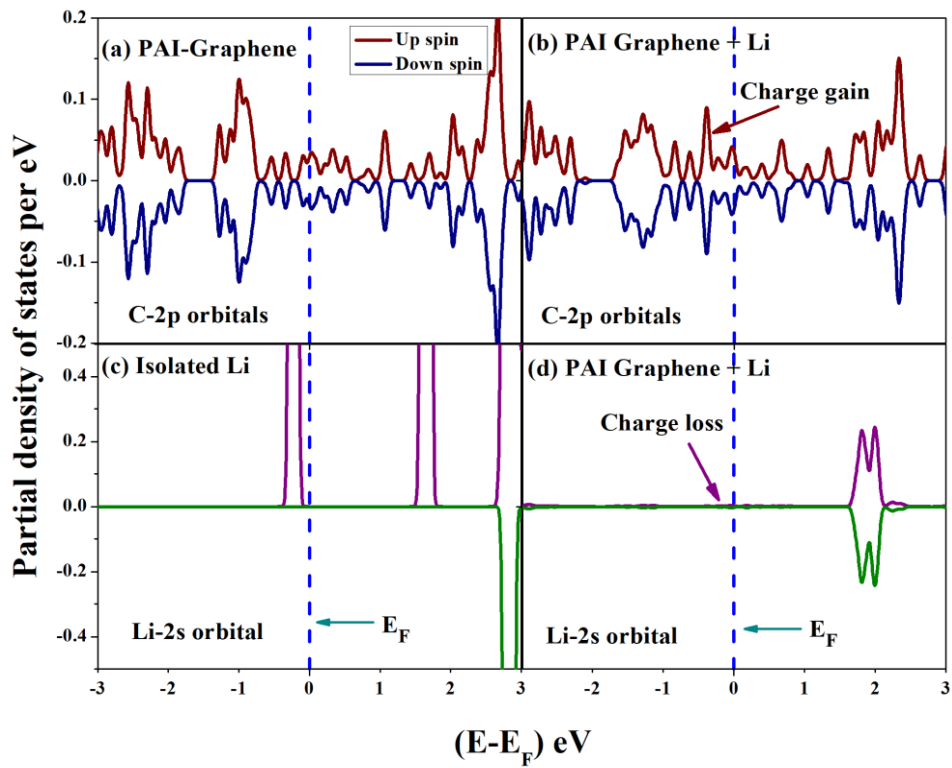

**Fig. 3. PDOS of (a) C-2p orbitals of PAI-graphene. (b) C-2p orbitals of Li-decorated PAI-graphene. (c) Li-2s orbitals of isolated Li atom. (d) Li-2s orbitals of Li-decorated PAI-graphene.**

Next, we present the surface plot for the charge density difference of the Li-decorated PAI-graphene system in **Fig. 4.** This plot is in R-G-B color code pattern, and with an isosurface value of 0.057 e, in which the red-colored regions denote the charge accumulation and the green-colored sphere denotes the Li-atom. The electronic charge of the 2s-orbitals of Li-atom has been transferred to the nearest carbon atoms (heptagon) of the PAI-graphene. The PDOS and charge density difference plots are consistent and explain the orbital interaction mechanism between metal atom Li and the host PAI-graphene structure.

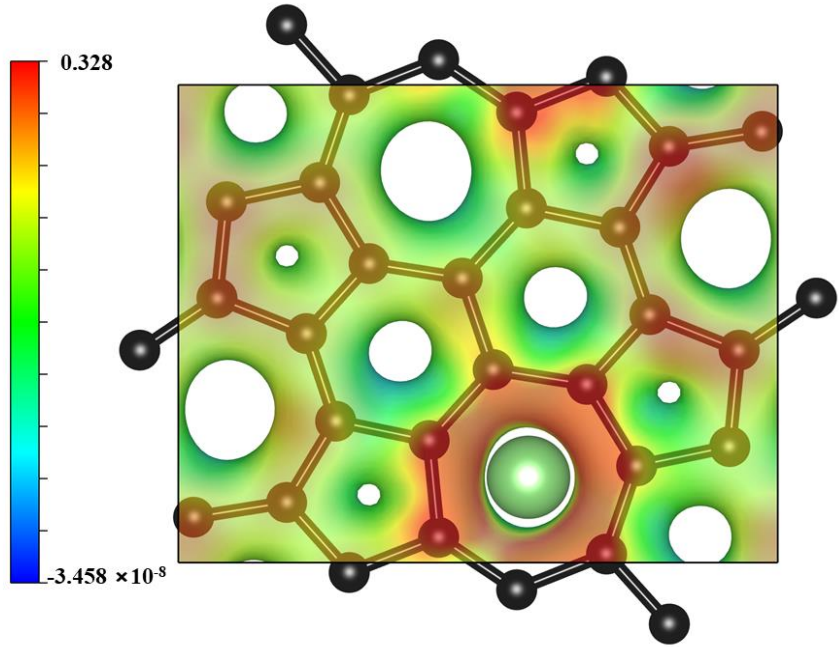

**Fig. 4. Charge density difference surface plot for PAI-graphene + Li and PAI-graphene ρ (PAI-graphene + Li) – ρ (PAI-graphene) systems. Here, iso-surface value is 0.057 e and plot is in according to R-G-B color pattern.**

**Reversible adsorption of $H_2$ molecules to Li-decorated PAI-graphene**

Next, we put the 1$^{st}$ $H_2$ on the top of the Li-decorated PAI-graphene and performed the relaxation calculations. The $H_2$ molecule is kept 2 Å distance far from the Li atom of the PAI-graphene + Li system. We have taken the heptagon position of the Li-atom for the Li-decorated PAI-graphene system because at this site, the Li atom is bounded with the maximum B.E. of -2 eV. The relaxed geometry of PAI-graphene + Li + 1 $H_2$ is presented in **Fig. 5 (a).** The bond length between Li and 1$^{st}$ $H_2$ is found to be 2.11 Å after the relaxation. We note that the difference in the H-H bond distance is negligible (~ 0.01 Å) compared to isolated $H_2$. For the adsorption energy calculations of the attached $H_2$ molecules, the zero-point energy and entropy correction effects are also accounted from the literature[83]. The adsorption energy of the 1$^{st}$ $H_2$ absorbed on the Li atom of Li-decorated PAI-graphene is found to be -0.29 eV. The hydrogen molecule is bounded by the electrostatic interaction acting between the Li and H-atoms of the $H_2$. Next, we put more hydrogen molecules successively and performed geometry optimization. The adsorption energy of the 2$^{nd}$, 3$^{rd}$, and 4$^{th}$ H$_2$ molecules are found to be -0.28 eV, -0.22 eV, and -0.19 eV, respectively. The B.E. of all four H$_2$ is close to the range -0.2 eV to -0.4 eV and is quite appropriate for the reversible attachment of H$_2$ molecules.[5]

The optimized structures of PAI-graphene + Li + 2 H$_2$, PAI-graphene + Li + 3 H$_2$, and PAI-graphene + Li + 4 H$_2$ compositions are presented in **Fig. 5 (b), Fig. 5 (c),** and, **Fig. 5 (d),** respectively.

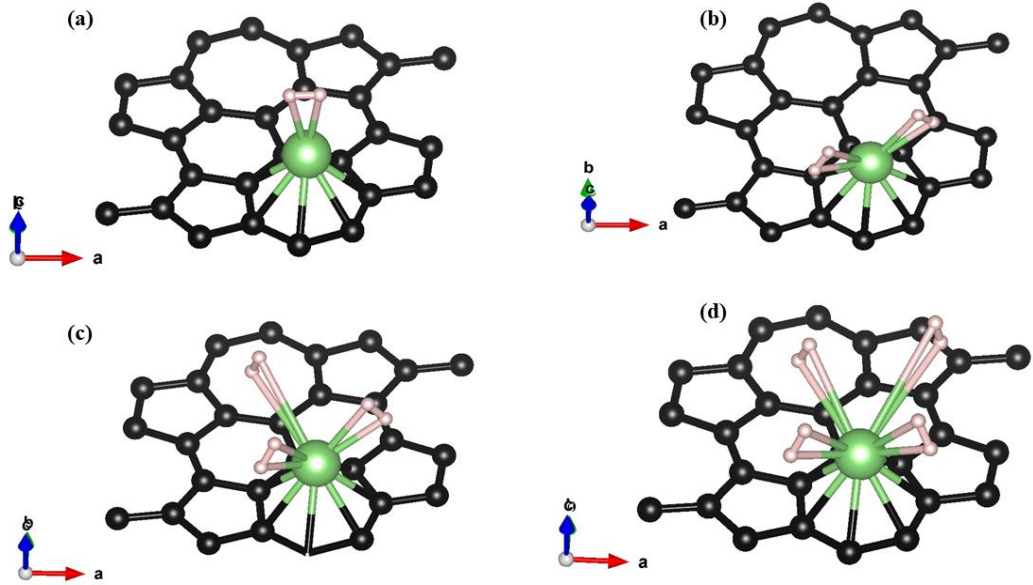

**Fig. 5. Relaxed geometries of (a) PAI-graphene + Li + H$_2$ (b) PAI-graphene + Li + 2H$_2$ (c) PAI-graphene + Li + 3H$_2$, and (d) PAI-graphene + Li + 4H$_2$ compositions. Black, green and light pink color sphere denote carbon, lithium and hydrogen atom, respectively.**

The adsorption energy of the n$^{th}$ hydrogen molecule adsorbed in the current step can be calculated by using the equation given below:

$$E_n (H_2) = E (PAI - gra + Li + n\, H_2) - E (PAI - gra + Li + (n-1)H_2) - E(H_2) \quad (2)$$

Here, $E (PAI - gra + Li + n\, H_2)$, and $E (PAI - gra + Li + (n-1)H_2)$ are the total energy of the n and (n-1) H$_2$ molecules attached to the Li-decorated PAI-graphene system, while $E(H_2)$ is the energy of the single isolated H$_2$. The adsorption energy of the H$_2$ molecules, desorption temperatures, average C-Li, and H-H distances for all the compositions are provided in **Table 1.**

**Table 1. Adsorption energies, desorption temperatures, and average bond distances (C-Li, Li-H, and H-H) for the PAI-graphene + Li + n H$_2$ (for n = 1 to 4) compositions.**

| Compositions PAI-graphene + Li + (n) H$_2$ n= 1 to 4 | Adsorption energy (eV) | Desorption temperature (K) | Average C-Li distance (Å) | Average Li-H distance (Å) | Average H-H distance (Å) |
|---|---|---|---|---|---|
| PAI-graphene + Li + H$_2$ | -0.29 | 374 | 2.24 | 2.11 | 0.754 |
| PAI-graphene + Li + 2 H$_2$ | -0.28 | 361 | 2.27 | 2.17 | 0.756 |
| PAI-graphene + Li + 3 H$_2$ | -0.22 | 284 | 2.27 | 2.48 | 0.756 |
| PAI-graphene + Li + 4 H$_2$ | -0.19 | 245 | 2.28 | 2.72 | 0.753 |

To explore the orbital interactions and charge flow direction, we present the PDOS of H-1s orbital of isolated H$_2$ and PAI-graphene + Li + 1 H$_2$ structures in **Fig. 6 (a & b).** We have noticed some enhancement in the PDOS near the Fermi level for the 1s-orbital of hydrogen when H$_2$ molecule is attached to Li-decorated PAI-graphene as displayed in **Fig. 6 (b),** implying that the charge is flowing from the Li-2s orbital to the H-1s orbital of H$_2$, when H$_2$ is attached to Li-decorated PAI-graphene due to the difference in the electronegativities of Li and H atoms.

Therefore, H$_2$ molecules are attached to the PAI-graphene + Li system by electrostatic interactions.

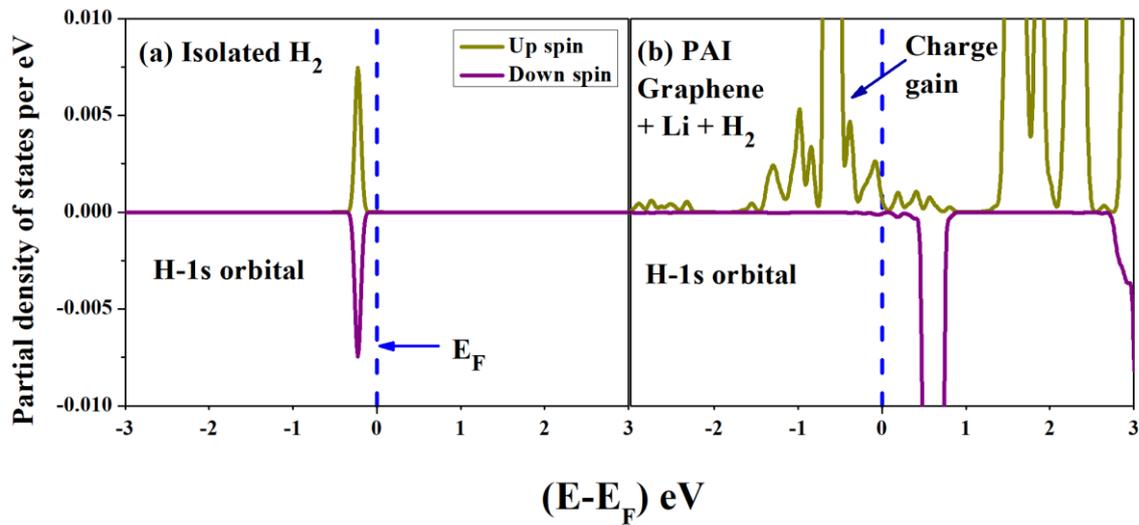

**Fig. 6. PDOS for (a) H-1s orbital of isolated H$_2$ molecule. (b) H-1s orbital for PAI-graphene + Li + H$_2$.**

Next, we estimated the desorption temperatures for the adsorbed H$_2$ molecules by applying Van't Hoff formula [84]:

$$T_d = \left(\frac{E_b}{k_B}\right)\left(\frac{\Delta S}{R} - \ln P/P_0\right)^{-1} \qquad (3)$$

Here, $T_d$ is the desorption temperature of the H$_2$ corresponding to the adsorption energy $E_b$. $\Delta S$ is the change in the entropy of H$_2$ while undergoing from gas to liquid state[85]. $P_0$, R, and $k_B$ are the atmospheric pressure, gas constant, and Boltzmann constant, respectively. P is equilibrium pressure, taken to be 1 atm for desorption temperature estimation. The desorption temperatures of the first, second, third, and fourth H$_2$ attached to the Li-decorated PAI-graphene system were found to be 374 K, 361 K, 284 K, and 245 K, respectively. The desorption temperature of individual H$_2$ is significantly higher than the boiling point of liquid nitrogen (77 K) and critical point of hydrogen (33K), which can be further enhanced by increasing the equilibrium pressure for room temperature hydrogen storage.

**Absence of metal clustering and hydrogen uptake**

For the hydrogen uptake estimation, we have investigated the metal-metal clustering issue in the Li-decorated PAI-graphene system. As mentioned earlier, the Li atom is bounded to the PAI-graphene structure with almost similar B.E. at penta, hexa, and hepta sites (~ 2 eV). Keeping this in mind, if we put the metal atom at all these sites, then the calculated weight percentage will be very high. But, that configuration might leads to the clustering of Li atoms and is not suitable for practical applications. Therefore, we put the metal atom only at the hepta sites of PAI-graphene and calculated the gravimetric weight percentage of hydrogen. A single unit cell of PAI-graphene can be decorated by 8 Li atoms, as displayed in **Fig. 7.**, and we note that each Li atom attached to the PAI-graphene can adsorb 4 $H_2$ molecules from the DFT simulations. The gravimetric weight percentage of hydrogen is 15.7 % for the metal loading pattern, as displayed in **Fig. 7**, which is remarkably higher than the DOE guidelines of 6.5 %. The gravimetric weight percentage of the Li-decorated PAI-graphene system is calculated by using the following equation:

$$H_2\ Weight\ \% = \frac{n \times M(H_2)}{M\ (PAI - graphene) + m \times M(Li) + n \times M(H_2)} \quad (4)$$

Where 'M' denotes the molecular mass of the composition. 'm' is the total number of Li atoms decorated on a single unit cell of PAI-graphene, and 'n' $H_2$ molecules attached to the total 'm' Li-atoms. Here, we have taken m = 8 and n = 4 × m = 32 since each Li atom can attach 4 $H_2$ molecules.

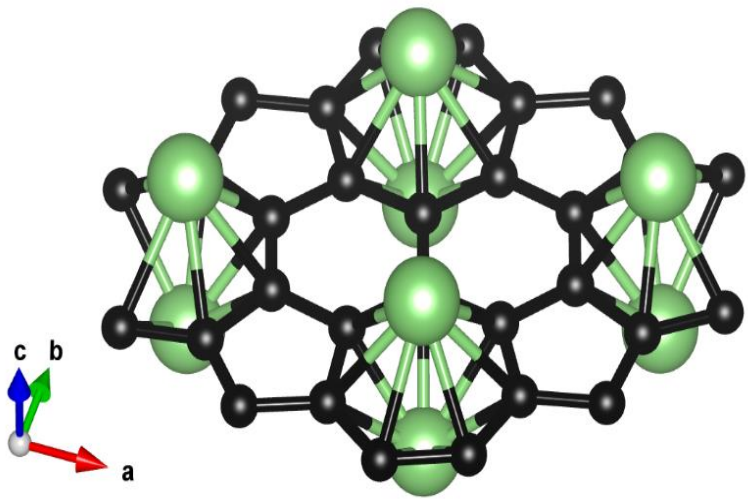

**Fig. 7. Metal loading pattern for the Li-atom to estimate the hydrogen uptake. One unit cell of PAI-graphene can adsorb 8 Li-atoms.**

Further, we investigate the possibilities of Li-atoms clustering in our system by calculating the energy barrier for the diffusion of Li atoms from one site (hepta) to the nearest hepta site of Li adsorption.

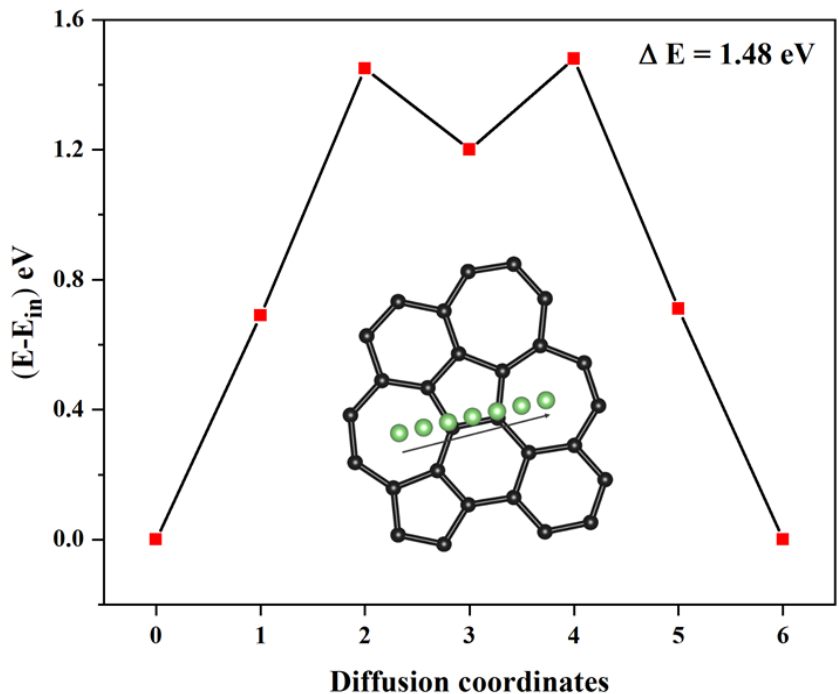

**Fig. 8. CI-NEB energy barrier plot for the Li atom from one stable site to the nearest similar site.**

The diffusion energy barrier is calculated by using the climbing-image nudged elastic band (CI-NEB) [80] method, in which the minimum energy pathway for the displacement of Li-atom is calculated within five equidistant transition states. The energy difference between a transition state and the initial state is plotted with respect to the diffusion coordinates as presented in **Fig. 8**. We found that an energy barrier of 1.48 eV is present in our system for the diffusion of Li atoms from one hepta site to the nearest hepta site.

## Conclusions

We have explored the reversible hydrogen storage properties of the Li-decorated carbon allotrope PAI-graphene by employing DFT simulations. Li atom is decorated to the top of the pentagon, hexagon, and heptagon of the PAI-graphene with almost same B.E. (~ 2 eV), indicating that a large number of metal atoms can be decorated on the PAI-graphene. We have investigated the interaction between the valence orbitals of Li and PAI-graphene by plotting DOS, PDOS, and charge density difference plots. It was found that one-unit cell of PAI-graphene can be decorated by 8 Li atoms, while each Li atom can adsorb 4 $H_2$ reversibly, leading to an ultrahigh hydrogen uptake of 15.7 % for the system. The binding energies and desorption temperatures of the $H_2$ molecules are optimum for the reversible use of hydrogen. The $H_2$ molecules are bounded to the Li atom by electrostatic interactions. Since the B.E. of the Li atom decorated on PAI-graphene, is more than the cohesive energy of the bulk Li, and an energy barrier of 1.48 eV is present for the diffusion of the Li atom, the Li-decorated PAI-graphene system will be prevented from metal clustering. Therefore, Li-decorated PAI-graphene is a potential hydrogen storage material with an ultrahigh gravimetric weight percentage (~16 %) of hydrogen.

## Acknowledgment

VM would like to acknowledge DST-INSPIRE for providing the fellowship and SpaceTime-2 supercomputing facility at IIT Bombay for the computing time. BC would like to thank Dr. T. Shakuntala and Dr. Nandini Garg for support and encouragement. BC also acknowledge support from Dr. S.M. Yusuf and Dr. A. K Mohanty.